# Comparison of Spreadsheets with other development tools (limitations, solutions, workarounds and alternatives)

Simon Murphy,
Codematic Limited,
Kinkry Hill House
Roadhead
Carlisle
UK
simon.murphy@codematic.net

## ABSTRACT

The spreadsheet paradigm has some unique risks and challenges that are not present in more traditional development technologies. Many of the recent advances in other branches of software development have bypassed spreadsheets and spreadsheet developers.

This paper compares spreadsheets and spreadsheet development to more traditional platforms such as databases and procedural languages.

It also considers the fundamental danger introduced in the transition from paper spreadsheets to electronic.

Suggestions are made to manage the risks and work around the limitations.

## 1. INTRODUCTION

**"400 Million users cant be wrong!"** (Microsoft, 2005)

Spreadsheet usage is almost universal (or endemic – depending on your view point). If cashflow is the lifeblood of business, spreadsheets are the language. They enjoy rather more widespread use than the paper original ever did. They are currently used for analysing, modelling, reporting and forecasting billions and billions of pounds worth of business transactions daily. There is no evidence of widespread business collapses due to spreadsheet errors (but the odd bankruptcy has been known).

Spreadsheets are being used for more and more ambitious projects, many are well beyond that envisaged by the original creators in the 70's. Are we beyond the limit? In the 70's 'Garbage In, Garbage Out' was the standard. In 2005 that is just not acceptable 'Garbage In, Error message out' or possibly even better 'No Garbage in' is the modern standard. (McConnell, 2004)

This paper compares spreadsheets to other development tools, looks at some of the problems associated with spreadsheets as a development platform and suggests workarounds and solutions.

## 2. COMPARISIONS WITH OTHER TOOLS

**Select by location not value.**

Paper spreadsheets have 1 mode of access, by the user, by value (you look down the title column looking for the text that describes the items you want). Electronic spreadsheets have 2 modes of access – the same user mode, by value, and the underlying, formula based, 'by location' access mode. This disconnect is unique to electronic spreadsheets and is a fundamental weakness that guarantees fragile systems.

This duality is a significant barrier to understanding and auditing non trivial spreadsheets.

Simple Demo:
Are the Gross Profit formulas correct?

|   | A | B | C | D | E | F |
|---|---|---|---|---|---|---|
| 1 | Simple Gross Profit calculation - enter +ve numbers | | | | | |
| 2 |   |   |   | 2003 | 2004 | 2005 |
| 3 |   | Sales |   |   |   |   |
| 4 |   |   |   |   |   |   |
| 5 |   | COGS |   |   |   |   |
| 6 |   |   |   |   |   |   |
| 7 |   | Gross Profit |   | =D3-D5 | =E3-E5 | =F3-F5 |
| 8 |   |   |   |   |   |   |

What about these?

|   |   | 2003 | 2004 | 2005 |
|---|---|---|---|---|
| Sales |   |   |   |   |
| COGS |   |   |   |   |
| Gross Profit |   | =R[-4]C-R[-2]C | =R[-4]C-R[-2]C | =R[-4]C-R[-2]C |

|   |   | 2003 | 2004 | 2005 |
|---|---|---|---|---|
| Sales |   |   |   |   |
| COGS |   |   |   |   |
| Gross Profit |   | =D3-D5 | =E3-E5 | =F3-F5 |

|   | A | B | C | D | E | F |
|---|---|---|---|---|---|---|
| 1 | Simple Gross Profit calculation - enter +ve numbers | | | | | |
| 2 |   |   |   | 2003 | 2004 | 2005 |
| 3 |   | Sales |   |   |   |   |
| 4 |   |   |   |   |   |   |
| 5 |   | COGS |   |   |   |   |
| 6 |   |   |   |   |   |   |
| 7 |   | Gross Profit |   | =Sales-COGS | =Sales-COGS | =Sales-COGS |
| 8 |   |   |   |   |   |   |

(Yes, No, No and No)

The actual (not apparent) spatial relationship is critical to understanding and testing a spreadsheet. The appearance and layout are irrelevant at best, and often downright misleading. Row and column headers must be visible to understand the model. The connection between meaningful labels and executable logic is coincidental.

Alternative development platforms such as databases rely on a select by value approach. As in 'SELECT * FROM PL WHERE LineItem = "Sales"'. This is much more robust. The human readable labels are used by the software.

**Type Safe**

Type safety is a contract that a program will not perform an operation on a variable that is not valid for that data type. Modern languages are more and more rigorously type safe. Most compilers will warn of an attempt to assign a string value to a numerical data type. Spreadsheets have no real comprehension of data types, you can put anything in any cell.

**Scope**

Modern programming best practice recommends minimising the visibility of variables. Block scope is preferred to routine, which is preferred to module which is preferred to global data. If an application really needs global data to function, this is a strong sign of significant design flaws (McConnell, 2004). In a spreadsheet, cells have global read visibility. Any other cell anywhere can see the value in any cell. This prevents the reliable use of information hiding and interface programming. Erwig suggests this global visibility puts spreadsheets in the same category as assembly language (Erwig, 2004).

**Data separation**

In N-Tier architectures there is the data tier, the business logic tier(s) and presentation tier. This separation allows each part to be optimised for its particular purpose, and minimises the effects of changes. In a spreadsheet everything is commonly lumped together, and presentation requirements often take priority over documenting complex business rules.

**Security**

Server based architectures are inherently more secure than desktop, and compiled binaries are difficult to modify maliciously or accidentally. Modern databases provide role based security that can be integrated with the operating system and applied at the record or field level. Worksheet protection is trivial to bypass, and often counter productive, workbook open protection is irrelevant if the user needs to open the workbook to use it.

**Scalability**

A program routine is written once and used many times, whereas each spreadsheet cell needs its own version of a formula. A VB program to take some numbers and add them is much more complex than a spreadsheet 'SUM()' formula. But the VB code to sum a thousand sets of numbers has the same complexity, a spreadsheet would have 1,000 formulas, each needing to be checked for correctness, arguably 1,000 times (or more) more complex.

**Development tools**

The Latest version of Visual Studio (VS 2005) assists the developer to create UML diagrams to represent the system, automatically generate database schemas and code, work with databases, write code to implement business rules, design the user interface (eg Web, windows forms, even Excel). It provides security and traceability for development resources through a source control system, it offers unit testing with automatic creation of test cases. All without leaving the development environment, all with context sensitive help and tips. Spreadsheets offer a few intrinsic tools to assist with development and testing but the difference in scope, power and flexibility is dramatic.

Panko suggests spreadsheet development is in a similar condition to mainstream development in the 60's (Panko, 1998) – and he's right, and so are the tools.

**VBA**

Commercial Excel VBA code is generally of appalling quality, most of it breaking every recommended best practice. The Excel/VBA link is not robust and appropriate use of named ranges to connect code and worksheet cells is rare, string constant references are more common.

**Ad-hoc**

Spreadsheets are a superbly powerful and flexible ad-hoc analysis and presentation tool for a single user. The second best tool for everything (Powell, 2004). Unfortunately ad-hoc tools lead to ad-hoc designs, ad-hoc designs are hard to test, hard to maintain, and hard to extend. With no formal development lifecycle or migration plan, models live on and develop beyond their initial life expectancy and scope.

**Links**

Inter-workbook links create hidden dependencies and make data consistency difficult to assess. Links enable circular references that Excel cannot spot, unless all linked workbooks are open at once.

Example - Analysis of link sources for 1 live, commercial workbook:

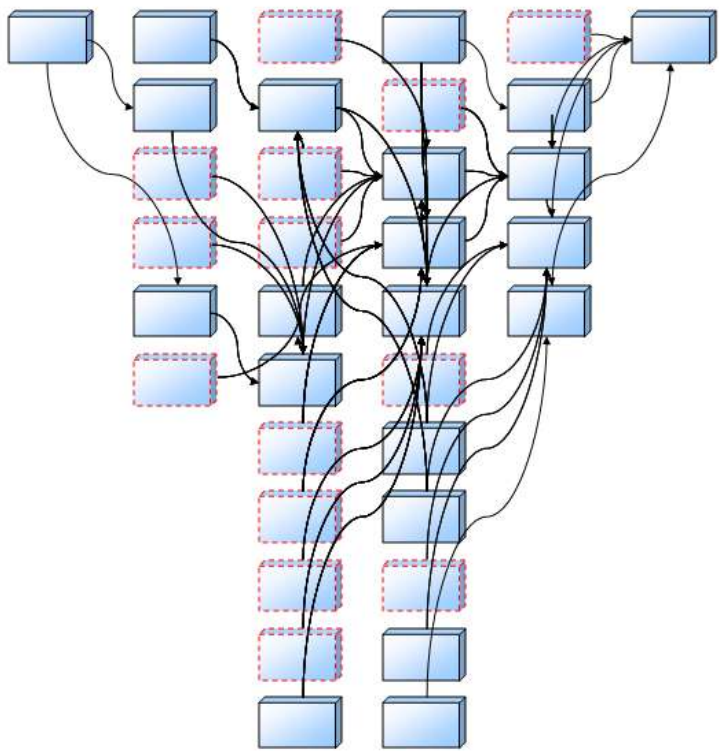

34 linked workbooks, 20 of which were found, 14 workbooks missing so unchecked for further links, over 100 links found. Chances of it being correct? Depends on your definition of correct: 100%(if you mean correct enough), 0%(if you mean totally, provably correct).

## 3.  SUGGESTIONS, SOLUTIONS AND WORKAROUNDS

**General**

The main advice is to be aware of the limitations of spreadsheets.

If you are a developer, then you owe it to your clients to know enough about alternative development platforms to be able to advise when a spreadsheet might not be the best choice.

If you are an end user you need to be aware of the signs your spreadsheet analysis may have outgrown its current implementation.  (eg: unwieldy, difficult to modify, difficult to reconcile, incomprehensible, only usable by the original author)

If you are a manager you need to be aware of where your information comes from and how it gets to you.  Teams of analysts working in Excel all day may not be creating value, they may be creating a monster that will eventually paralyse your business (the so called Spreadsheet Hell).  Warning signs are long delays in answering apparently simple questions, regular errors, reports that don't reconcile to other sources, limited team skills outside spreadsheets, aversion to working with each others models, lack of formal IT training, lack of IT department interaction.

| | Summary Table | | | |
|---|---|---|---|---|
| | Issue | Cause | Impact | Management suggestions |
| | | (apart from spreadsheet flexibility) | (apart from fragility) | (apart from use a more robust tool) |
| | | | | |
| 1 | Select by location not value | Visual approach to modelling | Spreadsheet view is disconnected from the user view | Focus spreadsheet efforts on small, short lived ad-hoc models. Keep dependent and similar items close together. Use label driven methods where correctness is important, such as consolidations from external sources (eg LOOKUP (4 argument version), MATCH, VBA) watch for performance degradation. Consider database (could still be in Excel) |
| 2 | Not type safe | Allows rapid modelling and quick changes | Visible representation of cell contents may be misleading | Use good visual design and layout to clarify the type of data required. Use data validation to control data entry, but beware of its limitations. Use code or forms to thoroughly validate input. If incorrect data types represent a major risk, use a strongly typed tool |
| 3 | Global Scope | Simple reuse of previous analysis | Inner workings cannot be hidden to allow later changes with no side effects | Design sheets with clear blocks and areas to highlight cells that may be used else where. |
| 4 | Lack of Data/Logic separation | Reduces need for forethought and design | Comprehension is reduced | Use appropriate layout to aid understanding. Large models should be broken down into simple blocks. |
| 5 | Lack of security | Primarily a single user tool | Intellectual property cannot be protected, spreadsheets can't be trusted once distributed | If non trivial security is required don't use spreadsheets. Before implementing any security be clear on what, and who the risks are, and what the real world impact of any likely breach is. |
| 6 | Poor scalability | lack of true data/logic separation means different data cannot be run through the same logic | exponential spreadsheet complexity v problem complexity relationship | Focus spreadsheet developments around the problem identification and solution evaluation phases rather than the implementation of a solution. |
| 7 | Poor development tools | Lack of user demand, tool builder complacency, and risky economics | Development time longer than need be, errors easy to add hard to find | Although Excel tools are somewhat limited there are many third party tools that pay for themselves in minutes |
| 8 | Poor quality VBA | Poor use of (freely and easily available) developer training | VBA is often more of a burden than an enabler | Make the effort to learn some of the industry best practices developed over 30-40 years to reduce complexity, improve quality and minimise risk of errors. |
| 9 | Ad-hoc nature of spreadsheets | Commercial pressure | Difficult to maintain, enhance and test. | Use large paper, or white boards to break out the elements of the model into shapes, fill in detail until what you are to build becomes clear. |
| 10 | Dangerous use of links | Quick reuse of previous results | Results may be inconsistent and or unrepeatable | Use a vba import routine with a date stamp and user name stamp. |

## 4. OTHER FACTORS

Many researchers propose extra tools, methodologies, or training to impart some structure and robustness into spreadsheets and spreadsheet use. They miss several key facts:
1. People use spreadsheets **because** of their flexibility, not in spite of it.
2. Most people already have a robust tool for building structured models on their desktops. It's called Microsoft Access, and most people ignore it because it's not flexible enough for them.
3. Behind every spreadsheet horror, there is a deadline driven manager who prioritizes information timeliness over accuracy. In the modern commercial world where competitive advantage can last minutes (or less), wrong information is better than no information (as long as it's not too wrong!)

If spreadsheets are so fragile and error prone why is so much work done with them? Cost, speed of development and current skill set. Excluding these 3 factors spreadsheets are probably never the right tool. But who can exclude these factors?

It has been suggested that spreadsheet use or abuse is an organisational thing (Cleary, 2004), commercial experience backs this up.

Spreadsheet use creates a web that quickly develops in uncontrolled environments (500 new spreadsheets per year (net of deletions), per analyst in one organisation (approx 2Gb of data suggesting an average size of 4Mb per model (typically 20-30 worksheets, 40,000 non blank cells per workbook) equates to 20 Million new data items per year (per analyst))).

Spreadsheets make a superb requirements development tool, and a great prototyping tool, but as every software textbook or developer will tell you, you must throw the prototype away. If you are building a racing car you wouldn't start with a go cart, but you may make a clay prototype to test the aerodynamics.

## 5. THE FUTURE

Spreadsheets are the new legacy system with many organisations managing down their reliance on spreadsheets. Total replacement projects have had limited success. Technical solutions do not fix cultural problems.

A stronger business school focus on commercial database use rather than spreadsheets would prepare students for the modern work world of data manipulation rather than creation. The amount of information available electronically now is worlds away from what was available even 5-10 years ago.

Microsoft has woken up to the problems and potential of Excel. Expect to see a lot of work in this area as Microsoft attempt to leverage their ownership of the corporate desktop.

## 6. CONCLUSION

Errors or quality must be related to 'fit for purpose', and many commercial spreadsheets are probably just about good enough. A surprisingly large margin of error would not be catastrophic in many models. This tolerance is demonstrated by the lack of wholesale collapse of spreadsheet addicted organisations. Spreadsheet models are only one of the sources of information available and other sources may carry more weight.

Bad spreadsheets are a symptom not a cause. To address the problems associated with spreadsheets, the culture of spreadsheet abuse must be addressed.

Spreadsheets are a superb tool with many valuable uses. They do have limits however, and are frequently misused and abused. Blaming spreadsheets for the commercial challenges they introduce makes as much sense as blaming cars for car crashes. There is the occasional mechanical problem, but far and away the biggest culprit is operator error.

If you use spreadsheets you should know their limits as well as your own. Enter the spreadsheet maze at your own risk and with your eyes open. You have a choice of tools, choose wisely.

**References**

Cleary P, (2004) IEEE Foundations of Spreadsheets Workshop

Erwig, M, (2004) IEEE Foundations of Spreadsheets Workshop

McConnell, S, (2004), Code Complete, Microsoft

Microsoft, www.microsoft.com/presspass/press/2003/oct03/10-13VSTOOfficeLaunchPR.asp – (accessed 16 May 2005)

Panko, R, (1998), "What we know about Spreadsheet Errors", Journal of End User Computing

Powell, SG and Baker KR, (2004), The Art of Modelling with Spreadsheets, Wiley